\documentclass{emulateapj}
\usepackage{apjfonts}
\usepackage{rotating}

\newcommand{\pivec}{\mbox{\boldmath $\pi$}}

\lefthead{SHIN ET AL.} 
\righthead{BROWN DWARF COMPANION MICROLENSING BINARIES}

\begin{document}
\title{MICROLENSING BINARIES WITH CANDIDATE BROWN DWARF COMPANIONS
}

\author{
I.-G. Shin$^{1,73}$,       
C. Han$^{1,73,78}$,
A. Gould$^{2,73}$, 
A. Udalski$^{3,71}$,    
T. Sumi$^{4,72}$,        
M. Dominik$^{5,74,75,76,77}$,
J.-P. Beaulieu$^{6,76}$,
Y. Tsapras$^{7,38,74}$,
V. Bozza$^{8,54,75}$,\\
and\\
M.\ K.\ Szyma{\'n}ski$^{3}$,
M.\ Kubiak$^{3}$, 
I.\ Soszy{\'n}ski$^{3}$,
G.\ Pietrzy{\'n}ski$^{3,9}$, 
R.\ Poleski$^{3}$, 
K.\ Ulaczyk$^{3}$, 
P.\ Pietrukowicz$^{3}$,
S.\ Koz{\l}owski$^{3}$,
J.\ Skowron$^{2,73}$,
{\L}.\ Wyrzykowski$^{3,10}$\\ 
(The OGLE Collaboration),\\
F. Abe$^{11}$,        
D.\ P.\ Bennett$^{12,76}$,  
I.\ A.\ Bond$^{13}$,
C.\ S.\ Botzler$^{14}$,
M.\ Freeman$^{14}$,   
A.\ Fukui$^{11}$,   
K.\ Furusawa$^{11}$,   
F.\ Hayashi$^{11}$,    
J.\ B.\ Hearnshaw$^{15}$, 
S.\ Hosaka$^{11}$,     
Y.\ Itow$^{11}$,        
K.\ Kamiya$^{11}$,      
P.\ M.\ Kilmartin$^{16}$, 
S.\ Kobara$^{11}$,
A.\ Korpela$^{17}$,     
W.\ Lin$^{13}$,         
C.\ H.\ Ling$^{13}$,     
S.\ Makita$^{11}$,      
K.\ Masuda$^{11}$,      
Y.\ Matsubara$^{11}$,    
N.\ Miyake$^{11}$,
Y.\ Muraki$^{18}$,     
M.\ Nagaya$^{11}$,
K.\ Nishimoto$^{11}$,   
K.\ Ohnishi$^{19}$,    
T.\ Okumura$^{11}$,
K.\ Omori$^{11}$,
Y.\ C.\ Perrott$^{14}$,   
N.\ Rattenbury$^{14}$,   
To.\ Saito$^{20}$,    
L.\ Skuljan$^{13}$,     
D.\ J.\ Sullivan$^{17}$,  
D.\ Suzuki$^{4}$,      
W.\ L.\ Sweatman$^{13}$, 
P.\ J.\ Tristram$^{16}$,  
K.\ Wada$^{4}$,        
P.\ C.\ M.\ Yock$^{14}$\\   
(The MOA Collaboration),\\
G.\ W.\ Christie$^{21}$,   
D.\ L.\ Depoy$^{22}$,      
S.\ Dong$^{23}$, 
A.\ Gal-Yam$^{24}$,
B.\ S.\ Gaudi$^{2}$,
L.-W. Hung$^{25}$,
J.\ Janczak$^{26}$,
S.\ Kaspi$^{27}$,
D.\ Maoz$^{27}$,
J.\ McCormick$^{28}$,
D.\ McGregor$^{2}$,    
D.\ Moorhouse$^{29}$,    
J.\ A.\ Mu\~{n}oz$^{30}$,
T.\ Natusch$^{21}$,
C.\ Nelson$^{31}$,
R.\ W.\ Pogge$^{2}$,
T.-G. TAN$^{32}$,      
D.\ Polishook$^{27}$,    
Y.\ Shvartzvald$^{27}$,
A.\ Shporer$^{7,70}$,
G.\ Thornley$^{29}$,
U.\ Malamud$^{27}$,
J.\ C.\ Yee$^{2}$,\\
and\\
J.-Y. Choi$^{1}$,
Y.-K. Jung$^{1}$,
H. Park$^{1}$,
C.-U. Lee$^{33}$,
B.-G. Park$^{33}$,
J.-R. Koo$^{33}$\\
(The $\mu$FUN Collaboration),\\
D.\ Bajek$^{5}$,
D.\ M.\ Bramich$^{34,75}$,
P.\ Browne$^{5,75}$, 
K.\ Horne$^{5,76}$,
S.\ Ipatov$^{35}$,          
C.\ Snodgrass$^{36,56,75}$,  
I.\ Steele$^{37}$,  
R.\ Street$^{7}$\\       
(The RoboNet Collaboration),\\
K.\ A.\ Alsubai$^{35,74}$,
M.\ J.\ Burgdorf$^{39}$,
S.\ Calchi Novati$^{8,40}$,
P.\ Dodds$^{5,74}$,
S.\ Dreizler$^{41}$,
X.-S. Fang$^{42}$,
F.\ Grundahl$^{43}$,
C.-H. Gu$^{42}$,
S.\ Hardis$^{44}$,
K.\ Harps{\o}e$^{44,69}$,
T.\ C.\ Hinse$^{33,44,51}$,
M.\ Hundertmark$^{5,41,74}$,
J.\ Jessen-Hansen$^{43}$,
U.\ G.\ J{\o}rgensen$^{44,69}$,
N.\ Kains$^{5,34,74}$,
E.\ Kerins$^{45}$,
C.\ Liebig$^{5,46,74}$,
M.\ Lund$^{43}$,
M.\ Lundkvist$^{43}$,
L.\ Mancini$^{8,47}$,
M.\ Mathiasen$^{44}$,
A.\ Hornstrup$^{48}$,
M.\ T.\ Penny$^{45}$,
S.\ Proft$^{46}$,
S.\ Rahvar$^{49,38,55}$,
D.\ Ricci$^{50}$,
G.\ Scarpetta$^{8,52}$,
J.\ Skottfelt$^{44}$,
J.\ Southworth$^{53}$,
J.\ Surdej$^{50}$,
J.\ Tregloan-Reed$^{53}$,
O.\ Wertz$^{50}$,
F.\ Zimmer$^{46}$\\ 
(The MiNDSTEp Consortium),\\
M.\ D.\ Albrow$^{15}$,
V.\ Batista$^{2,73}$,    
S.\ Brillant$^{56}$,     
J.\ A.\ R.\ Caldwell$^{57}$,
J.\ J.\ Calitz$^{58}$,
A.\ Cassan$^{6}$,       
A.\ Cole$^{59}$,         
K.\ H.\ Cook$^{60}$,
E.\ Corrales$^{6}$,      
Ch.\ Coutures$^{6}$,     
S.\ Dieters$^{6,61}$,    
D.\ Dominis Prester$^{62}$,
J.\ Donatowicz$^{63}$,
P.\ Fouqu\'e$^{61}$,     
J.\ Greenhill$^{59}$,
K.\ Hill$^{59}$,     
M.\ Hoffman$^{58}$,
S.\ R. Kane$^{64}$,
D.\ Kubas$^{6,56}$,
J.-B. Marquette$^{6}$,
R.\ Martin$^{65}$,
P.\ Meintjes$^{58}$,
J.\ Menzies$^{66}$,
K.\ R.\ Pollard$^{15}$, 
K.\ C.\ Sahu$^{67,75}$,
J.\ Wambsganss$^{68}$,
A.\ Williams$^{65,75}$,
C.\ Vinter$^{44}$,
M.\ Zub$^{68}$\\
(The PLANET Collaboration),\\
}

\bigskip\bigskip
\affil{$^{1}$Department of Physics, Institute for Astrophysics, Chungbuk National University, Cheongju 371-763, Republic of Korea}
\affil{$^{2}$Department of Astronomy, Ohio State University, 140 W. 18th Ave., Columbus, OH 43210, USA}
\affil{$^{3}$Warsaw University Observatory, Al. Ujazdowskie 4, 00-478 Warszawa, Poland}
\affil{$^{4}$Department of Earth and Space Science, Osaka University, Osaka 560-0043, Japan}
\affil{$^{5}$SUPA, University of St. Andrews, School of Physics \& Astronomy, North Haugh, St. Andrews, KY16 9SS, United Kingdom}
\affil{$^{6}$Institut d'Astrophysique de Paris, UMR7095 CNRS--Universit{\'e} Pierre \& Marie Curie, 98 bis boulevard Arago, 75014 Paris, France}
\affil{$^{7}$Las Cumbres Observatory Global Telescope Network, 6740B Cortona Dr, Suite 102, Goleta, CA 93117, USA}
\affil{$^{8}$Universit\`{a} degli Studi di Salerno, Dipartimento di Fisica ``E.R. Caianiello'', Via S. Allende, 84081 Baronissi (SA), Italy}
\affil{$^{9}$Universidad de Concepci\'{o}n, Departamento de Astronomia, Casilla 160--C, Concepci\'{o}n, Chile}
\affil{$^{10}$Institute of Astronomy, University of Cambridge, Madingley Road, Cambridge CB3 0HA, United Kingdom}
\affil{$^{11}$Solar-Terrestrial Environment Laboratory, Nagoya University, Nagoya, 464-8601, Japan}
\affil{$^{12}$Department of Physics, University of Notre Damey, Notre Dame, IN 46556, USA}
\affil{$^{13}$Institute of Information and Mathematical Sciences, Massey University, Private Bag 102-904, North Shore Mail Centre, Auckland, New Zealand} 
\affil{$^{14}$Department of Physics, University of Auckland, Private Bag 92019, Auckland, New Zealand}
\affil{$^{15}$University of Canterbury, Department of Physics and Astronomy, Private Bag 4800, Christchurch 8020, New Zealand}  
\affil{$^{16}$Mt. John Observatory, P.O. Box 56, Lake Tekapo 8770, New Zealand} 
\affil{$^{17}$School of Chemical and Physical Sciences, Victoria University, Wellington, New Zealand} 
\affil{$^{18}$Department of Physics, Konan University, Nishiokamoto 8-9-1, Kobe 658-8501, Japan} 
\affil{$^{19}$Nagano National College of Technology, Nagano 381-8550, Japan} 
\affil{$^{20}$Tokyo Metropolitan College of Industrial Technology, Tokyo 116-8523, Japan} 
\affil{$^{21}$Auckland Observatory, P.O. Box 24-180, Auckland, New Zealand} 
\affil{$^{22}$Department of Physics, Texas A\&M University, College Station, TX, USA} 
\affil{$^{23}$Institute for Advanced Study, Einstein Drive, Princeton, NJ 08540, USA}
\affil{$^{24}$Benoziyo Center for Astrophysics, the Weizmann Institute, Israel} 
\affil{$^{25}$Department of Physics \& Astronomy, University of California Los Angeles, Los Angeles, CA 90095, USA}
\affil{$^{26}$Department of Physics, Ohio State University, 191 W. Woodruff, Columbus, OH 43210, USA}
\affil{$^{27}$School of Physics and Astronomy, Tel-Aviv University, Tel Aviv 69978, Israel} 
\affil{$^{28}$Farm Cove Observatory, Pakuranga, Auckland} 
\affil{$^{29}$Kumeu Observatory, Kumeu, New Zealand} 
\affil{$^{30}$Departamento de Astronomi{\'a} y Astrof{\'i}sica, Universidad de Valencia, E-46100 Burjassot, Valencia, Spain}
\affil{$^{31}$College of Optical Sciences, University of Arizona, 1630 E. University Blvd, Tucson Arizona, 85721, USA}
\affil{$^{32}$Perth Exoplanet Survey Telescope, Perth, Australia}
\affil{$^{33}$Korea Astronomy and Space Science Institute, 776 Daedukdae-ro, Yuseong-gu, Daejeon 305-348, Republic of Korea} 
\affil{$^{34}$European Southern Observatory, Karl-Schwarzschild-Stra{\ss}e 2, 85748 Garching bei M{\"u}nchen, Germany}
\affil{$^{35}$Qatar Foundation, P.O. Box 5825, Doha, Qatar}
\affil{$^{36}$Max-Planck-Institut f{\"o}r Sonnensystemforschung, Max-Planck-Str. 2, 37191 Katlenburg-Lindau, Germany}
\affil{$^{37}$Astrophysics Research Institute, Liverpool John Moores University, Egerton Wharf, Birkenhead CH41 1LD, United Kingdom}
\affil{$^{38}$School of Physics and Astronomy, Queen Mary, University of London, Mile End Road, London, E1 4NS, United Kingdom} 
\affil{$^{39}$HE Space Operations GmbH Flughafenallee 24 D-28199 Bremen, Germany}
\affil{$^{40}$Istituto Internazionale per gli Alti Studi Scientifici (IIASS), Vietri Sul Mare (SA), Italy}
\affil{$^{41}$Institut f\"{u}r Astrophysik, Georg-August-Universit\"{a}t, Friedrich-Hund-Platz 1, 37077 G\"{o}ttingen, Germany}
\affil{$^{42}$National Astronomical Observatories/Yunnan Observatory, Joint Laboratory for Optical Astronomy, Chinese Academy of Sciences, Kunming 650011, People's Republic of China }
\affil{$^{43}$Department of Physics \& Astronomy, Aarhus Universitet, Ny Munkegade, 8000 {\AA}rhus C, Denmark}
\affil{$^{44}$Niels Bohr Institute, University of Copenhagen, Juliane Maries vej 30, 2100 Copenhagen, Denmark}
\affil{$^{45}$Jodrell Bank Centre for Astrophysics, University of Manchester, Oxford Road,Manchester, M13 9PL, United Kingdom}
\affil{$^{46}$Astronomisches Rechen-Institut, Zentrum f\"{u}r Astronomie der Universit\"{a}t Heidelberg (ZAH),  M\"{o}nchhofstr.\ 12-14, 69120 Heidelberg, Germany}
\affil{$^{47}$Max Planck Institute for Astronomy, K\"{o}nigstuhl 17, 69117 Heidelberg, Germany}
\affil{$^{48}$Danmarks Tekniske Universitet, Institut for Rumforskning og-teknologi, Juliane Maries Vej 30, 2100 K{\o}benhavn, Denmark}
\affil{$^{49}$Department of Physics, Sharif University of Technology, P.~O.\ Box 11155--9161, Tehran, Iran}
\affil{$^{50}$Institut d'Astrophysique et de G\'{e}ophysique, All\'{e}e du 6 Ao\^{u}t 17, Sart Tilman, B\^{a}t.\ B5c, 4000 Li\`{e}ge, Belgium}
\affil{$^{51}$Armagh Observatory, College Hill, Armagh, BT61 9DG, Northern Ireland, United Kingdom}
\affil{$^{52}$INFN, Gruppo Collegato di Salerno, Sezione di Napoli, Italy}
\affil{$^{53}$Astrophysics Group, Keele University, Staffordshire, ST5 5BG, United Kingdom}
\affil{$^{54}$Istituto Nazionale di Fisica Nucleare, Sezione di Napoli, Italy}
\affil{$^{55}$Perimeter Institute for Theoretical Physics, 31 Caroline St. N., Waterloo ON, N2L 2Y5, Canada}
\affil{$^{56}$European Southern Observatory, Casilla 19001, Vitacura 19, Santiago, Chile} 
\affil{$^{57}$McDonald Observatory, 16120 St Hwy Spur 78 \#2, Fort Davis, TX 79334, USA}
\affil{$^{58}$University of the Free State, Faculty of Natural and Agricultural Sciences, Department of Physics, PO Box 339, Bloemfontein 9300, South Africa}
\affil{$^{59}$School of Math and Physics, University of Tasmania, Private Bag 37, GPO Hobart, Tasmania 7001, Australia} 
\affil{$^{60}$Institute of Geophysics and Planetary Physics (IGPP), L-413, Lawrence Livermore National Laboratory, PO Box 808, Livermore, CA 94451, USA}
\affil{$^{61}$LATT, Universit\'e de Toulouse, CNRS, 14 Avenue Edouard Belin, 31400 Toulouse, France} 
\affil{$^{62}$Physics Department, Faculty of Arts and Sciences, University of Rijeka, Omladinska 14, 51000 Rijeka, Croatia}
\affil{$^{63}$Technical University of Vienna, Department of Computing, Wiedner Hauptstrasse 10, Vienna, Austria}
\affil{$^{64}$NASA Exoplanet Science Institute, Caltech, MS 100-22, 770 South Wilson Avenue, Pasadena, CA 91125, USA}
\affil{$^{65}$Perth Observatory, Walnut Road, Bickley, Perth 6076, Australia}
\affil{$^{66}$South African Astronomical Observatory, P.O. Box 9 Observatory 7935, South Africa} 
\affil{$^{67}$Space Telescope Science Institute, 3700 San Martin Drive, Baltimore, MD 21218, USA}
\affil{$^{68}$Astronomisches Rechen-Institut (ARI), Zentrum f{\"u}r Astronomie der Universit{\"a}t Heidelberg (ZAH), M{\"o}nchhofstrasse 12-14, 69120 Heidelberg, Germany}
\affil{$^{69}$Centre for Star and Planet Formation, Geological Museum, {\O}ster Voldgade 5, 1350 Copenhagen, Denmark}
\affil{$^{70}$Department of Physics, University of California, Santa Barbara, CA 93106, USA}
\affil{$^{71}$The OGLE Collaboration}
\affil{$^{72}$The MOA Collaboration}
\affil{$^{73}$The $\mu$FUN Collaboration}
\affil{$^{74}$The RoboNet Collaboration}
\affil{$^{75}$The MiNDSTEp Consortium}
\affil{$^{76}$The PLANET Collaboration}
\affil{$^{77}$Royal Society University Research Fellow}
\affil{$^{78}$Corresponding author}

\begin{abstract}

Brown dwarfs are important objects because they may provide a missing link between stars and planets, two populations that have dramatically 
different formation history. In this paper, we present the candidate binaries with brown dwarf companions that are found by analyzing binary 
microlensing events discovered during 2004 -- 2011 observation seasons. Based on the low mass ratio criterion of $q < 0.2$, we found 7 candidate 
events, including OGLE-2004-BLG-035, OGLE-2004-BLG-039, OGLE-2007-BLG-006, OGLE-2007-BLG-399/MOA-2007-BLG-334, MOA-2011-BLG-104/OGLE-2011-BLG-0172, 
MOA-2011-BLG-149, and MOA-201-BLG-278/OGLE-2011-BLG-012N. Among them, we are able to confirm that the companions of the lenses of MOA-2011-BLG-104/OGLE-2011-BLG-0172 
and MOA-2011-BLG-149 are brown dwarfs by determining the mass of the lens based on the simultaneous measurement of the Einstein radius and the lens 
parallax. The measured mass of the brown dwarf companions are $(0.02\pm0.01)$ $M_{\odot}$ and $(0.019\pm0.002)$ $M_{\odot}$ for MOA-2011-BLG-104/OGLE-2011-BLG-0172 
and MOA-2011-BLG-149, respectively, and both companions are orbiting low mass M dwarf host stars. More microlensing brown dwarfs are expected to 
be detected as the number of lensing events with well covered light curves increases with new generation searches.

\end{abstract}


\section{INTRODUCTION}

Brown dwarfs are sub-stellar objects that are too low in mass to sustain hydrogen fusion reactions in their cores but much higher than planets. 
Studies of brown dwarfs are important because they may provide a missing link between stars and planets \citep{kulkarni97}, two populations that 
have dramatically different formation history. In addition, the Galaxy may be teeming with brown dwarfs although there is no evidence for a large 
population of brown dwarfs from current observations \citep{graff96, najita00, tisserand07}. Brown dwarfs had long been thought to exist based 
on theoretical considerations \citep{kumar69}. However, these objects are intrinsically faint and thus the first confirmed brown dwarf was not 
discovered until 1995 \citep{rebolo95, nakajima95}.

There have been 2 major methods of detecting brown dwarfs. The first is direct imaging by using ground (e.g.,\ Schuh 2003) and space-borne 
(e.g.,\ Mainzer 2011) infrared instruments. As in the case of extrasolar planets, brown dwarfs can also be indirectly discovered by detecting 
wobbles in the motion of the companion star or the flux decrease of the companion star caused by the brown dwarf occulting the companion star.

Microlensing is also an effective method to detect brown dwarfs. Gravitational microlensing refers to the astronomical phenomenon wherein the 
brightness of a star is magnified by the bending of light caused by the gravity of an intervening object (lens) between the background star (source) 
and an observer. Since the phenomenon occurs regardless of the lens brightness, microlensing was proposed to detect dark components of the Galaxy such 
as black holes, neutron stars, and brown dwarfs \citep{paczynski86}.

Although effective, the application of the microlensing method in brown dwarf detections has been limited. The main reason for this is the 
difficulty in distinguishing brown dwarf events from those produced by main sequence stars. For general lensing events in which a single mass causes the 
brightening of a background star, the magnification of the lensed star flux depends on the projected lens-source separation, $u$, by
\begin{equation}
A={ u^2+2 \over { u(u^2+4)^{1/2} } },
\end{equation}
where the lens-source separation is normalized by the radius of the Einstein ring, $\theta_{\rm E}$ (Einstein radius). The lens-source separation is related to the 
lensing parameters of the time scale for the source to cross the Einstein radius, $t_{\rm E}$ (Einstein time scale), the time of the closest lens-source 
approach, $t_0$, and the lens-source separation at the moment of the closest approach, $u_0$ (impact parameter), by
\begin{equation}
u=\left[{ \left({t-t_0 \over{t_{\rm E}}}\right)^2 + {u_0}^2 }\right]^{1/2}.
\end{equation}
Among these lensing parameters, only the Einstein time scale provides information about the physical parameters of the lens. However, it results from 
the combination of the lens mass, distance, and transverse speed of the relative lens-source motion and thus it is difficult to uniquely determine the 
mass of the lensing object. Not knowing the mass, then, it is difficult to single out brown dwarf events from those produced 
by stars. In principle, it is possible to measure the lens mass by additionally measuring the Einstein radius and the lens parallax, which are respectively 
related to the physical parameters of the lens by
\begin{equation}
\theta_{\rm E} = \left( \kappa M \pi_{\rm rel} \right)^{1/2}
\end{equation}
and
\begin{equation}
\pi_{\rm E} ={ \pi_{\rm rel} \over{\theta_{\rm E} } },
\end{equation}
where $\pi_{\rm rel}={\rm AU}(D_{\rm L}^{-1}-D_{\rm S}^{-1})$, $\kappa=4G/(c^2{\rm AU})$, $M$ is the mass of the lens, and $D_{\rm L}$ and $D_{\rm S}$ are the 
distances to the lens and source star, respectively. The Einstein radius is measured by analyzing the distortion of the lensing light curve affected by the finite 
size of the source star \citep{nemiroff94, witt94, gould94}, while the lens parallax is measured by analyzing the deviation in a lensing 
light curve caused by the deviation of the relative lens-source motion from a rectilinear one due to the change of the observer's position induced by the orbital 
motion of the Earth around the Sun \citep{gould92}. Unfortunately, the chance of simultaneously measuring $\theta_{\rm E}$ and 
$\pi_{\rm E}$ is very low for a single-lens event. The finite source effect is important only for very high magnification events, for which the lens-source separation 
is comparable to the source size \citep{choi12}, but these events are very rare. The parallax effect is important only for events with time scales that are a significant fraction of 
the orbital period of the Earth, i.e.\ 1 yr. However, such long time-scale events are unlikely to be produced by low mass brown dwarfs because $t_{\rm E}\propto M^{1/2}$. As a result, 
there exist only two cases for which the lens of a single lensing event was identified as a brown dwarf \citep{smith03, gould09}
\footnote{Among these two cases, the microlensing brown dwarf discovered by \citet{gould09} is exception that provides the rule because the lens mass could be determined not by 
measuring the lens parallax induced by the Earth's orbital motion but from the ``terrestrial parallax''.}.

In binary lensing events, in which the lens is composed of two masses, on the other hand, the chance to identify the lens as a brown dwarf is relatively high. There are several reasons 
for this. First, for binary lensing events, it is possible to routinely measure the mass ratio, $q$, of the lens components. Then, brown dwarf 
candidates can be sorted out based on the first order criterion of small mass ratios. Second, the majority of binary lensing events involves caustic crossings \citep{schneider86, mao91}. 
The caustics represent the positions on the source plane at which the lensing magnification of a point source becomes infinite. The caustic-crossing part of a lensing light curve varies 
depending on the source size. Analysis of the light curve affected by this finite source effect allows one to measure the Einstein radius, which allows for better constraint on the lens mass. 
Third, the chance to measure the lens parallax is high, too. This is because the duration of an event corresponds to the total mass of the binary lens system not the low mass component 
and thus the average time scale of a binary lens event is longer than that of a single lens event.

Despite these advantages, there exists only one event (OGLE-2008-BLG-510) for which the companion of a binary lens is known to be a brown dwarf candidate \citep{bozza12}. The main reason for the rarity 
of brown dwarf events is the difficulty in analyzing binary lensing light curves due to the complexity of the light curves. For binary lensing events in which the mass ratio between the 
lens components is very low as in the case of a binary composed of a star and a planet, the signature of the low mass companion appears as a short term deviation on the top 
of an otherwise smooth and symmetric single lensing light curve of the primary. As a result, identification and analysis of planetary events are relatively simple, resulting in 
more than 20 identified microlensing planets. However, the perturbative nature of the companion breaks down and the light curve becomes very complex for binaries with mass ratios 
greater than $q\sim0.1$, which corresponds to the mass ratio of a typical binary lens composed of a brown dwarf and a star. This introduces difficulties in the immediate identification 
of a brown dwarf event. A firm identification of a brown dwarf event requires detailed modeling of lensing light curves. Unfortunately, modeling light curves of binary lensing events is a 
difficult task due to the complexity of $\chi^2$ surface in the parameter space combined with the heavy computation required for analysis. As a result, detailed modeling of binary 
events was conducted for a small subset of events, resulting in a small number of identified brown dwarfs. In recent years, however, there has been great progress in the analysis of binary 
events with the development of analysis methods based on advanced logic combined with improved computing power. This progress made the analysis not only more precise but also 
faster. As a result, nearly all binary lensing events are routinely modeled in real time with the progress of events in current microlensing experiments \citep{ryu10, bozza12}.

In this paper, we present brown dwarf candidates found from the analyses of binary lensing events detected from microlensing experiments conducted from 2004 to 2011. In \S 2, we describe 
the criteria of choosing brown dwarf candidates, observation of events, and data reduction. In \S 3, we describe the procedure of modeling. In \S 4, we present the results of the 
analysis. In \S 5, we draw our conclusions.

\begin{deluxetable*}{lrrrr}
\tablecaption{Coordinates of Events\label{table:one}}
\tablewidth{0pt}
\tablehead{
\multicolumn{1}{c}{event} &
\multicolumn{1}{c}{RA (J2000)} &
\multicolumn{1}{c}{DEC (J2000)} &
\multicolumn{1}{c}{$l$} &
\multicolumn{1}{c}{$b$}
}
\startdata
OGLE-2004-BLG-035                   & 17$^{\rm h}$48$^{\rm m}$43$^{\rm s}$\hskip-2pt.16 & -35$^\circ$57${\rm '}$45${\rm ''}$\hskip-2pt.9 & 354.32$^\circ$ & -4.19$^\circ$ \\
OGLE-2004-BLG-039                   & 17$^{\rm h}$53$^{\rm m}$47$^{\rm s}$\hskip-2pt.58 & -30$^\circ$52${\rm '}$11${\rm ''}$\hskip-2pt.7 & 359.25$^\circ$ & -2.51$^\circ$ \\
OGLE-2007-BLG-006                   & 18$^{\rm h}$02$^{\rm m}$52$^{\rm s}$\hskip-2pt.50 & -29$^\circ$15${\rm '}$11${\rm ''}$\hskip-2pt.8 &   1.63$^\circ$ & -3.41$^\circ$ \\
OGLE-2007-BLG-399/MOA-2007-BLG-334  & 17$^{\rm h}$45$^{\rm m}$35$^{\rm s}$\hskip-2pt.79 & -34$^\circ$54${\rm '}$37${\rm ''}$\hskip-2pt.5 & 354.89$^\circ$ & -3.10$^\circ$ \\
MOA-2011-BLG-104/OGLE-2011-BLG-0172 & 17$^{\rm h}$54$^{\rm m}$22$^{\rm s}$\hskip-2pt.48 & -29$^\circ$50${\rm '}$01${\rm ''}$\hskip-2pt.7 &   0.21$^\circ$ & -2.10$^\circ$ \\
MOA-2011-BLG-149                    & 17$^{\rm h}$56$^{\rm m}$47$^{\rm s}$\hskip-2pt.69 & -31$^\circ$16${\rm '}$04${\rm ''}$\hskip-2pt.7 & 359.23$^\circ$ & -3.27$^\circ$ \\
MOA-2011-BLG-278/OGLE-2011-BLG-012N & 17$^{\rm h}$54$^{\rm m}$11$^{\rm s}$\hskip-2pt.32 & -30$^\circ$05${\rm '}$21${\rm ''}$\hskip-2pt.6 & 359.97$^\circ$ & -2.19$^\circ$ 
\enddata  
\end{deluxetable*}

\begin{deluxetable}{ll}
\tablecaption{Telescopes\label{table:two}}
\tablewidth{0pt}
\tablehead{
\multicolumn{1}{c}{event} &
\multicolumn{1}{c}{telescopes}
}
\startdata
OGLE-2004-BLG-035   & OGLE, 1.3 m Warsaw, Las Campanas, Chile  \\
\hline
OGLE-2004-BLG-039   & OGLE, 1.3 m Warsaw, Las Campanas, Chile  \\ 
\hline
OGLE-2007-BLG-006   & OGLE, 1.3 m Warsaw, Las Campanas, Chile  \\ 
                    & $\mu$FUN, 1.3 m SMARTS, CTIO, Chile      \\
                    & $\mu$FUN, 0.4 m Auckland, New Zealand    \\
                    & $\mu$FUN, 0.4 m FCO, New Zealand         \\
\hline
OGLE-2007-BLG-399   & OGLE, 1.3 m Warsaw, Las Campanas, Chile  \\
/MOA-2007-BLG-334   & MOA, 2.0 m Mt. John, New Zealand         \\
                    & $\mu$FUN, 1.3 m SMARTS, CTIO, Chile      \\
                    & PLANET, 1.0 m SAAO, South Africa         \\
\hline
MOA-2011-BLG-104    & OGLE, 1.3 m Warsaw, Las Campanas, Chile  \\
/OGLE-2011-BLG-0172 & MOA, 2.0 m Mt. John, New Zealand         \\
                    & $\mu$FUN, 1.3 m SMARTS, CTIO, Chile      \\
                    & $\mu$FUN, 0.4 m Auckland, New Zealand    \\
                    & $\mu$FUN, 0.4 m FCO, New Zealand         \\
                    & $\mu$FUN, 0.5 m Wise, Israel             \\
                    & $\mu$FUN, 0.3 m PEST, Australia         \\
                    & RoboNet,  2.0 m FTN, Hawaii              \\
                    & RoboNet,  2.0 m FTS, Australia           \\
\hline
MOA-2011-BLG-149    & MOA, 2.0 m Mt. John, New Zealand         \\ 
                    & OGLE, 1.3 m Warsaw, Las Campanas, Chile  \\
                    & $\mu$FUN, 1.3 m SMARTS, CTIO, Chile      \\
                    & $\mu$FUN, 0.4 m Auckland, New Zealand    \\
                    & RoboNet, 2.0 m FTS, Australia            \\
                    & MiNDSTEp, 1.54 m Danish, La Silla, Chile \\
\hline
MOA-2011-BLG-278    & MOA, 2.0 m Mt. John, New Zealand         \\
/OGLE-2011-BLG-012N & OGLE, 1.3 m Warsaw, Las Campanas, Chile  \\ 
                    & $\mu$FUN, 1.3 m SMARTS, CTIO, Chile      \\
                    & $\mu$FUN, 0.4 m Auckland, New Zealand    \\
                    & $\mu$FUN, 0.4 m FCO, New Zealand         \\
                    & $\mu$FUN, 0.4 m Kumeu, New Zealand       \\
                    & $\mu$FUN, 1.0 m Lemmon, Arizona          \\
                    & RoboNet, 2.0 m FTS, Australia            \\
                    & MiNDSTEp, 1.54 m Danish, La Silla, Chile 
\enddata  
\tablecomments{
CTIO: Cerro Tololo Inter-American Observatory;
FCO: Farm Cove Observatory;
SAAO: South Africa Astronomical Observatory;
PEST: Perth Extrasolar Survey Telescope;
FTN: Faulkes North;
FTS: Faulkes South
}
\end{deluxetable}

\section{SAMPLE and OBSERVATIONS}

To search for brown dwarf candidates, we investigate binary lensing events discovered during 2004 -- 2011 microlensing observation seasons. Considering that typical Galactic microlensing 
events are produced by low mass stars \citep{han03}, the lower mass companion of an event with a mass ratio $q < 0.2$ is likely to be a brown dwarf candidate. Therefore, we sort out brown dwarf 
candidate events with the criterion of $q < 0.2$. For this, we conduct modeling of all binary lensing events with relatively good light curve coverage combined with obvious binary lensing features 
such as the spikes caused by crossings over the fold or approaches to the cusp of a caustic. The good coverage criterion is needed to secure the mass ratio measurement. The criterion 
of the characteristic features is needed because it is known that binary lensing events with weak signals can often be interpreted as other types of anomalies such as binary sources \citep{jaroszynski08}. 
For binary lensing events detected before 2008, we also refer to the previous analyses conducted by \citet{jaroszynski06}, \citet{skowron07}, and \citet{jaroszynski10}. From the 2004 -- 2011 search, we found 
7 candidate events, including OGLE-2004-BLG-035, OGLE-2004-BLG-039, OGLE-2007-BLG-006, OGLE-2007-BLG-399/MOA-2007-BLG-334, MOA-2011-BLG-104/OGLE-2011-BLG-0172, MOA-2011-BLG-149, and MOA-201-BLG-278/OGLE-2011-BLG-012N.

\begin{figure}[ht]
\epsscale{1.1}
\plotone{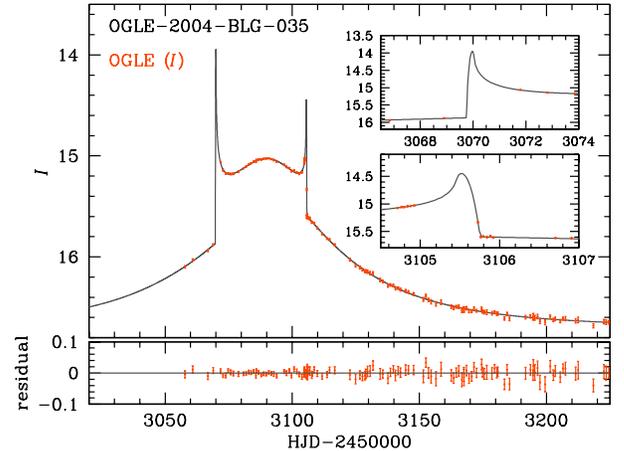}
\caption{\label{fig:one}
Light curve of OGLE-2004-BLG-035. The notation in the parentheses after the label of the observatory 
denotes the pass band of observation. The insets show the enlargement of the light curve during caustic 
crossings and approaches.  
}\end{figure}

\begin{figure}[ht]
\epsscale{1.1}
\plotone{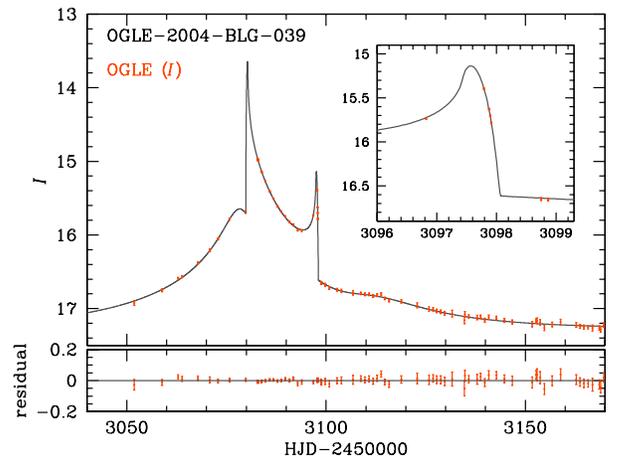}
\caption{\label{fig:two}
Light curve of OGLE-2004-BLG-039. Notations are same as in Fig \ref{fig:one}.
}\end{figure}

\begin{figure}[ht]
\epsscale{1.1}
\plotone{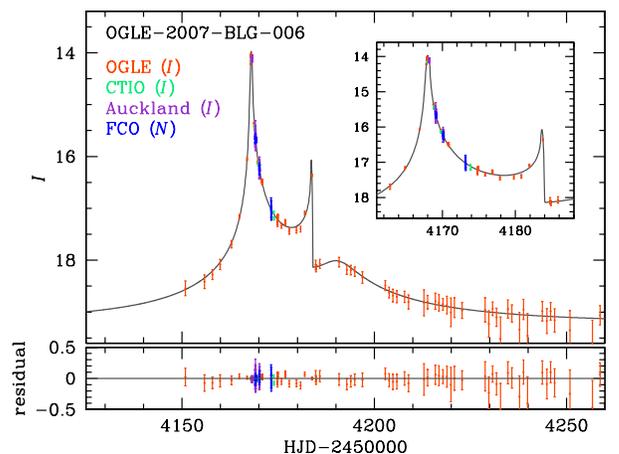}
\caption{\label{fig:three}
Light curve of OGLE-2007-BLG-006. Notations are same as in Fig \ref{fig:one}.
The notation for the observed passband ``$N$'' denotes no filter was used.
}\end{figure}

In Table \ref{table:one}, we list the candidate events that are analyzed in this work along with the equatorial and Galactic coordinates of the lensed stars. The name of each lensing event 
is formed by taking the survey group who discovered the event (i.e.\ OGLE or MOA), followed by the year of discovery (e.g.\ 2004), the direction of the observation field ("BLG" designating 
the Galactic bulge field), and a sequential number assigned to the event. If an event is discovered independently by two different groups, they are named separately. Survey observations 
were conducted by the two groups: the Optical Gravitational Lensing Experiment (OGLE: Udalski 2003) and the Microlensing Observation in Astrophysics (MOA: Bond et al. 2001, Sumi et al. 2003). Among 
7 events, 5 were additionally observed by follow-up observation groups including $\mu$FUN \citep{gould06}, PLANET \citep{beaulieu06}, RoboNet \citep{tsapras09}, and MiNDSTEp \citep{dominik10}. 
In Table \ref{table:two}, we list the survey and follow-up groups who participated in the observation of the individual events along with the telescopes and their locations.

Photometric reductions were carried out using photometry codes developed by the individual groups. The OGLE and MOA data were reduced by the photometry codes developed by \citet{udalski03} and \citet{bond01}, 
respectively, which are based on the Difference Image Analysis method \citep{alard98}. The $\mu$FUN data were processed using a pipeline based on the DoPHOT software \citep{schechter93}. For PLANET and 
MiNDSTEp data, a pipeline based on the pySIS software \citep{albrow09} is used. For the reduction of RoboNet data, the DanDIA pipeline \citep{bramich08} is used. Photometry errors of the individual data sets 
were rescaled so that $\chi^2$ per degree of freedom becomes unity for the data set of each observatory, where $\chi^2$ is estimated based on the best fit model. We eliminate data points with large errors and 
those lying beyond $3\sigma$ from the best fit model to minimize their effect on modeling. In Figures 1 -- 7, we present the light curves of the events.

\section{MODELING}

In the analyses of each lensing light curve, we search for the solution of lensing parameters that yield the best fit to the observed light curve. The basic structure of a binary lensing 
light curve is described by 7 parameters. The first three parameters $t_0$, $u_0$, and $t_{\rm E}$ are identical to those of a single-lensing event and they specify the source motion with respect to the lens. 
The next 3 parameters describe the binary lens system, including the projected separation between lens components in units of the Einstein radius, $s$, the mass ratio between the lens components, $q$, 
and the angle between source trajectory and the binary axis, $\alpha$. The final parameter is the source radius, $\theta_{\star}$, normalized to the Einstein radius, $\rho_{\star}=\theta_{\star}/\theta_{\rm E}$ 
(normalized source radius). The normalized source radius is needed to account for dilution of the lensing magnification caused by the finite size of the source. The finite source effect is important 
when the source is located at a position where the magnification varies very rapidly such as near the caustic and thus different parts of the source star are magnified by different amounts. In addition to 
these basic binary lensing parameters, it is often needed to consider the parallax effect for a fraction of events with long time scales. To describe the lens parallax effect, it is required to include the two 
additional parameters, $\pi_{{\rm E},N}$ and $\pi_{{\rm E},E}$, which represent the two components of the lens parallax vector $\pivec_{\rm E}$ projected on the sky along the north and east equatorial 
coordinates, respectively.  The size of the lens parallax vector corresponds to the ratio of the Earth's orbit to the Einstein radius projected on the observer observer's plane, 
i.e.\ $\pi_{\rm E}={\rm AU}/[r_{\rm E}D_{\rm S}/(D_{\rm S}-D_{\rm L})]$, where $r_{\rm E}=D_{\rm L}\theta_{\rm E}$ is the physical size of the Einstein radius. See Equation (4).

\begin{figure}[ht]
\epsscale{1.1}
\plotone{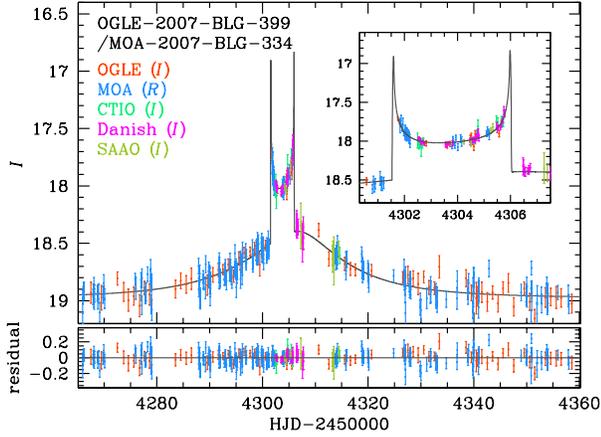}
\caption{\label{fig:four}
Light curve of OGLE-2007-BLG-399/MOA-2007-BLG-334. Notations are same as in Fig \ref{fig:one}.
}\end{figure}

\begin{figure}[ht]
\epsscale{1.1}
\plotone{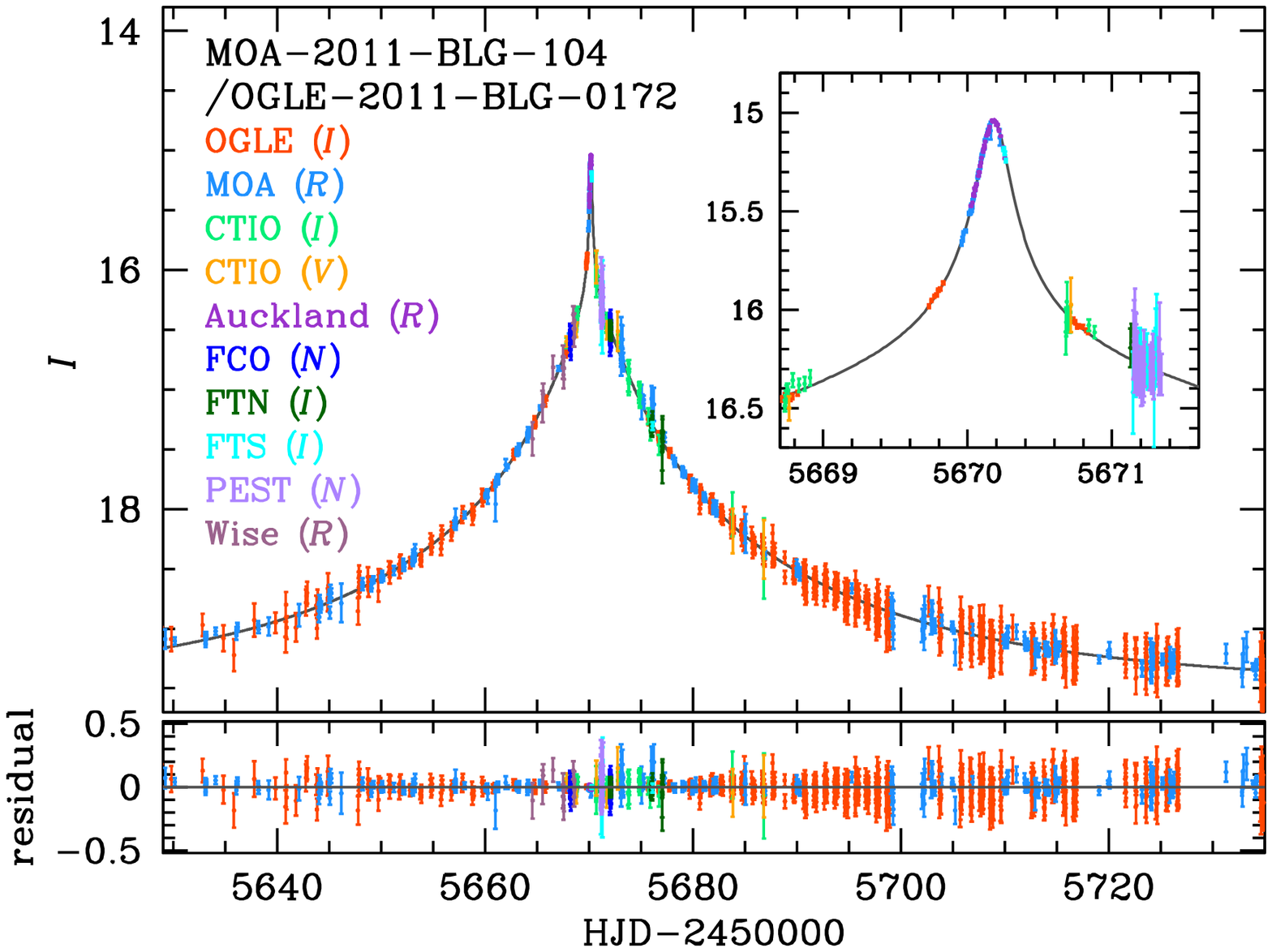}
\caption{\label{fig:five}
Light curve of MOA-2011-BLG-104/OGLE-2011-BLG-0172. Notations are same as in Fig \ref{fig:three}.
}\end{figure}

\begin{figure}[ht]
\epsscale{1.1}
\plotone{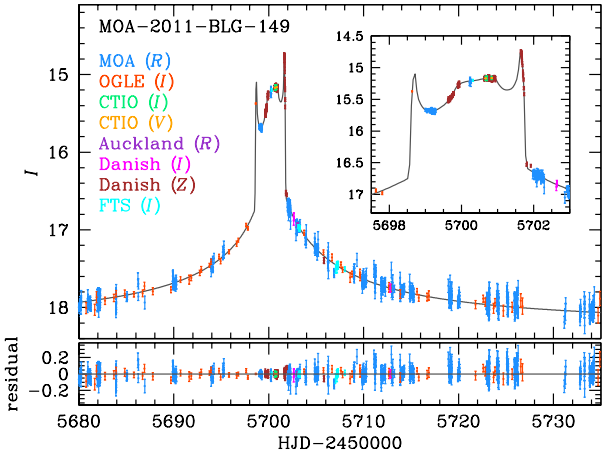}
\caption{\label{fig:six}
Light curve of MOA-2011-BLG-149. Notations are same as in Fig \ref{fig:one}.
}\end{figure}

\begin{figure}[ht]
\epsscale{1.1}
\plotone{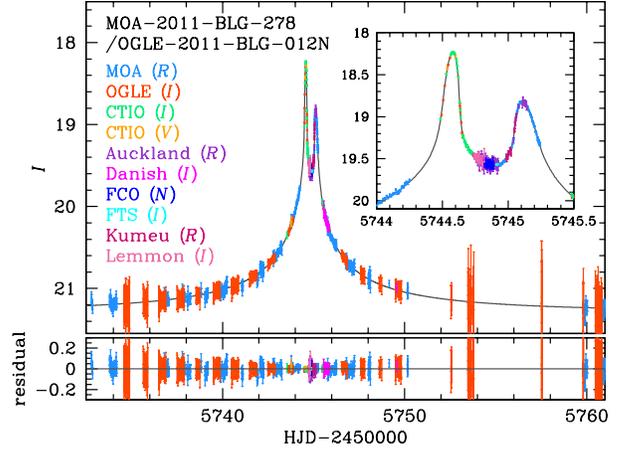}
\caption{\label{fig:seven}
Light curve of MOA-2011-BLG-278/OGLE-2011-BLG-012N. Notations are same as in Fig \ref{fig:three}.
}\end{figure}

In the analysis, we search for the best fit solution in two steps. In the first step, grid search is conducted for a subset of parameters while other parameters are optimized by using a downhill approach 
\citep{dong06}. By inspecting the $\chi^2$ distribution in the grid parameter space, we identify local minima. In the second step, we investigate the individual local minima and further refine the solutions by 
allowing all parameters to vary. Once the solutions corresponding to the local minima are found, we choose the best fit solution by comparing the $\chi^2$ values of the individual local minima. In this process, we also 
identify possible degenerate solutions resulting from various causes. We set $s$, $q$, and $\alpha$ as grid parameters because these parameters are related to light curve features in a complex way such that a 
small change in the values of the parameters can lead to dramatic changes in the resulting light curve. For the $\chi^2$ minimization in the downhill approach, we use the Markov Chain Monte Carlo (MCMC) method. 
Once the solution of the parameters is found, we estimate the uncertainties of the individual parameters based on the distributions of the parameters obtained from the MCMC chain of solutions.

\begin{figure*}[ht]
\epsscale{1.1}
\plotone{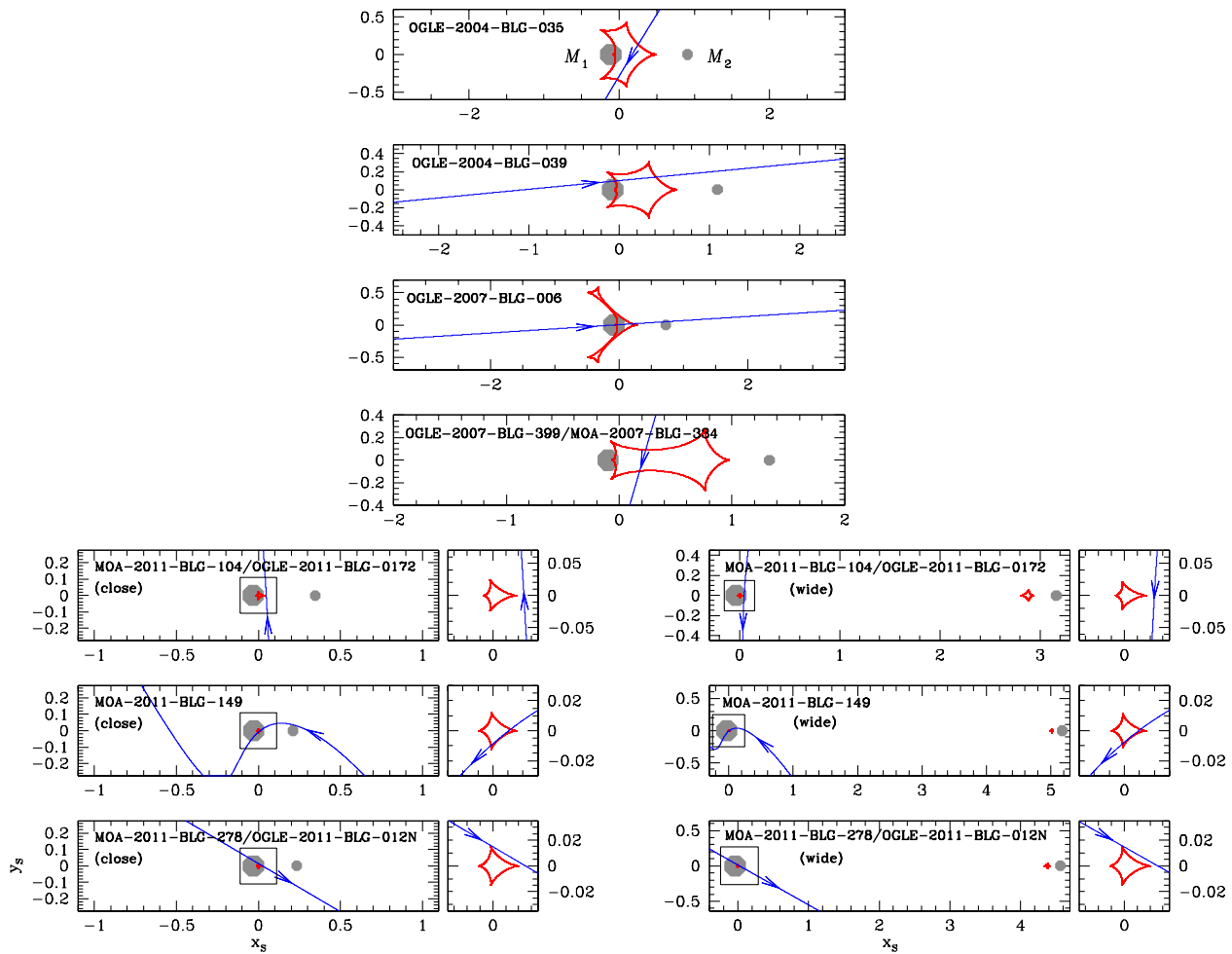}
\caption{\label{fig:eight}
Geometry of the binary lens systems for the light curves presented in Fig.\ 1 -- 7. In each panel, the two dots 
(marked by $M_1$ and $M_2$) represent the locations of the binary lens components, where the bigger ($M_1$) and 
smaller ($M_2$) dots represent the higher and lower mass binary components, respectively. The figure composed of 
concave curves is the caustic. The line with an arrow represents the source trajectory. For three events of 
MOA-2011-BLG-104/OGLE-2011-BLG-0172, MOA-2011-BLG-149 and MOA-2011-BLG-278/OGLE-2011-BLG-012N, we present the 
geometry of two solutions caused by the close/wide degeneracy. To better show the caustic-involved perturbations 
region for each of the three events with degenerate close and wide solutions, we present the enlargement of the 
perturbation region (enclosed by a box) in a separate panel on the right side of each main panel.  
}\end{figure*}

We use the inverse ray-shooting technique \citep{schneider86, kayser86, wambsganss97} to compute magnifications of events affected by the finite source effect. In this numerical method, rays are uniformly shot from the 
image plane, bent by the lens equation, and land on the source plane. Then, the finite magnification is computed as the ratio of the number density of rays on the source plane to that on the image plane. The lens equation 
for a binary lens, which describes the relation between the positions of a ray on the image and source planes, is represented by
\begin{equation}
\zeta=z-{m_1 \over{\bar{z}-\bar{z}_{\rm L,1}}}-{m_2 \over{\bar{z}-\bar{z}_{\rm L,2}}},
\end{equation}
where $m_1=1/(1+q)$ and $m_2=qm_1$ are the mass fractions of the binary lens components, $\zeta=\xi+i\eta$, $z=x+iy$, and $z_{\rm L}=x_{\rm L}+iy_{\rm L}$ represent the complex notations of the source, image, and lens 
locations, respectively, and $\bar{z}$ denotes the complex conjugate of $z$. The ray-shooting method requires heavy computation. 
We thus minimize computation time by applying the ray-shooting method only when the source crosses or approaches very close to caustics and utilizing the semi-analytic hexadecapole approximation \citep{gould08, pejcha09} 
in the adjacent area.

In computing finite magnifications, we consider the variation of the magnification caused by the limb-darkening of the source star surface for events with well-resolved features involving caustics. In our model, 
the surface brightness profile is described by
\begin{equation}
S_{\lambda}={F_{\lambda} \over{\pi{\theta_{\star}}^2}} \left[ 1-\Gamma_{\lambda} \left( 1- {3\over{2}} \cos\psi  \right) \right],
\end{equation}
where $\Gamma_{\lambda}$ is the linear limb-darkening coefficient, $F_{\lambda}$ is the source star flux, and $\psi$ is the angle between the normal to the source star's surface and the line of sight toward the star. 
We set the limb-darkening coefficients corresponding to the source type that are determined based on the color and magnitude of the source. In Table \ref{table:three}, we present the adopted limb-darkening coefficients, 
corresponding source types, and the measured de-reddened color along with the assumed values of the effective temperature, $T_{\rm eff}$, the surface turbulence velocity, $v_{\rm turb}$, and the surface gravity, $\log{g}$.

For all 7 events, we conduct modeling based on the 7 basic binary lensing parameters. We carry out additional modeling considering the parallax effect for events with Einstein time scales $t_{\rm E}>30$ days. 
When the parallax effect is considered, we separately investigate solutions with impact parameters $u_0>0$ and $u_0<0$. This is because the lensing light curves resulting from the source trajectories with $+u_0$ and $-u_0$ 
are different due to the asymmetry of the trajectory caused by the parallax effect, while the light curves are identical when the trajectories are straight lines due to the symmetry of the magnification pattern with 
respect to the binary axis.

\section{RESULTS}

\begin{deluxetable*}{cccc}
\tablecaption{Limb-darkening coefficients and source information\label{table:three}}
\tablewidth{0pt}
\tablehead{
\multicolumn{1}{l}{quantity} &
\multicolumn{1}{c}{MOA-2011-BLG-104} &
\multicolumn{1}{c}{MOA-2011-BLG-149} &
\multicolumn{1}{c}{MOA-2011-BLG-278} \\
\multicolumn{1}{l}{} &
\multicolumn{1}{c}{/OGLE-2011-BLG-0172} &
\multicolumn{1}{l}{} &
\multicolumn{1}{c}{/OGLE-2011-BLG-012N} 
}
\startdata
$\Gamma_{V}$                       & 0.51 & 0.73 & 0.59 \\
$\Gamma_{R}$                       & 0.43 & 0.64 & 0.51 \\
$\Gamma_{I}$                       & 0.36 & 0.53 & 0.43 \\
$(V-I)_0$                          & 0.62 & 1.64 & 0.85 \\
source type                        & FV   & KV   & GV   \\
$T_{\rm eff}$ ~($\rm K$)           & 6650 & 4410 & 5790 \\
$v_{\rm turb}$ ~($\rm km\ s^{-1}$) & 2    & 2    & 2    \\
$\log{g}$ ~($\rm cm\ s^{-2}$)      & 4.5  & 4.5  & 4.5  
\enddata  
\end{deluxetable*}

\begin{deluxetable*}{lrrrrrrrrrrrr}
\tablecaption{Model Parameters\label{table:four}}
\tablewidth{0pt}
\tablehead{
\multicolumn{1}{c}{event} &
\multicolumn{1}{c}{$\chi^2/{\rm dof}$} &
\multicolumn{1}{r}{$t_0$ (HJD')} &
\multicolumn{1}{c}{$u_0$} &
\multicolumn{1}{c}{$t_{\rm E}$ (days)} &
\multicolumn{1}{c}{$s$} &
\multicolumn{1}{c}{$q$} &
\multicolumn{1}{c}{$\alpha$} &
\multicolumn{1}{c}{$\rho_\star$ ($10^{-3}$)} &
\multicolumn{1}{c}{$\pi_{{\rm E},N}$} &
\multicolumn{1}{c}{$\pi_{{\rm E},E}$} 
}
\startdata
OGLE-2004-BLG-035 & 270.8/240  & 3085.825   &  0.1502      & 66.81     & 1.021      & 0.122      &  2.129      & 2.36      & --        & --        \\
                  &            & $\pm$0.090 &  $\pm$0.0011 & $\pm$0.29 & $\pm$0.002 & $\pm$0.002 &  $\pm$0.014 & $\pm$0.16 & --        & --        \\
                  & 246.6/240  & 3085.651   &  0.1475      & 65.00     & 1.024      & 0.130      &  2.113      & 2.24      & -0.02     & 0.07      \\
                  &            & $\pm$0.167 &  $\pm$0.0017 & $\pm$0.68 & $\pm$0.003 & $\pm$0.003 &  $\pm$0.019 & $\pm$0.27 & $\pm$0.06 & $\pm$0.02 \\
\hline
OGLE-2004-BLG-039 & 393.0/386  & 3081.422   &  0.1009      & 38.95     & 1.163      & 0.093      &  6.188      & 6.64      &  --       &  --       \\
                  &            & $\pm$0.054 &  $\pm$0.0031 & $\pm$0.67 & $\pm$0.005 & $\pm$0.004 &  $\pm$0.008 & $\pm$0.39 &  --       &  --       \\
\hline
OGLE-2007-BLG-006 & 325.6/587  & 4171.109   &  0.0043      & 60.92     & 0.801      & 0.110      &  6.219      & 1.70      &  --       &  --       \\
                  &            & $\pm$0.161 &  $\pm$0.0003 & $\pm$3.53 & $\pm$0.014 & $\pm$0.009 &  $\pm$0.004 & $\pm$0.15 &  --       &  --       \\
\hline
OGLE-2007-BLG-399  & 1289.8/999 & 4305.037   &  0.1906      & 23.39     & 1.446      & 0.168      &  1.855      &  --       &  --       &  --       \\
/MOA-2007-BLG-334  &            & $\pm$0.129 &  $\pm$0.0244 & $\pm$2.23 & $\pm$0.047 & $\pm$0.016 &  $\pm$0.017 &  --       &  --       &  --       \\
\hline
MOA-2011-BLG-104    & 1726.9/3210 & 5670.304   & 0.0478      & 38.86     & 0.379      & 0.089      & 1.638      & 1.50      &  --       &  --       \\
/OGLE-2011-BLG-0172 &             & $\pm$0.005 & $\pm$0.0008 & $\pm$0.48 & $\pm$0.004 & $\pm$0.003 & $\pm$0.003 & $\pm$0.12 &  --       &  --       \\
                    & 1724.6/3210 & 5670.310   & 0.0478      & 40.06     & 3.213      & 0.118      & 1.639      & 1.36      &  --       &  --       \\
                    &             & $\pm$0.005 & $\pm$0.0007 & $\pm$0.48 & $\pm$0.048 & $\pm$0.005 & $\pm$0.003 & $\pm$0.13 &  --       &  --       \\
                    & 1712.4/3208 & 5670.286   & -0.0481     & 38.47     & 0.379      & 0.090      & -1.628     & 1.54      & -0.19     & 0.10      \\
                    &             & $\pm$0.006 & $\pm$0.0009 & $\pm$0.49 & $\pm$0.005 & $\pm$0.003 & $\pm$0.003 & $\pm$0.13 & $\pm$0.30 & $\pm$0.03 \\
                    & 1709.9/3208 & 5670.291   & 0.0494      & 39.29     & 3.195      & 0.121      & 1.628      & 1.36      & -0.33     & 0.08      \\
                    &             & $\pm$0.006 & $\pm$0.0008 & $\pm$0.52 & $\pm$0.048 & $\pm$0.005 & $\pm$0.003 & $\pm$0.13 & $\pm$0.21 & $\pm$0.03 \\
\hline
MOA-2011-BLG-149    & 5031.1/4918 & 5700.402   & 0.0081      & 148.60    & 0.256      & 0.145      & 2.450      & 0.55      &  --       &  --        \\
                    &             & $\pm$0.008 & $\pm$0.0003 & $\pm$4.97 & $\pm$0.003 & $\pm$0.005 & $\pm$0.004 & $\pm$0.02 &  --       &  --        \\
                    & 5036.1/4918 & 5700.447   & 0.0080      & 150.60    & 5.077      & 0.229      & 2.464      & 0.55      &  --       &  --        \\
                    &             & $\pm$0.008 & $\pm$0.0003 & $\pm$5.74 & $\pm$0.082 & $\pm$0.013 & $\pm$0.004 & $\pm$0.02 &  --       &  --        \\
                    & 4889.0/4916 & 5700.399   & 0.0067      & 179.73    & 0.240      & 0.134      & 2.446      & 0.44      & -0.74     & -0.24      \\
                    &             & $\pm$0.008 & $\pm$0.0003 & $\pm$8.50 & $\pm$0.004 & $\pm$0.007 & $\pm$0.004 & $\pm$0.02 & $\pm$0.04 & $\pm$0.02  \\
                    & 4892.1/4916 & 5700.441   & 0.0064      & 183.91    & 5.210      & 0.190      & 2.460      & 0.44      & -0.69     & -0.20      \\
                    &             & $\pm$0.008 & $\pm$0.0002 & $\pm$5.66 & $\pm$0.088 & $\pm$0.012 & $\pm$0.003 & $\pm$0.02 & $\pm$0.04 & $\pm$0.02  \\
\hline
MOA-2011-BLG-278    & 5820.7/5558 & 5744.741   & 0.0128      & 18.96     & 0.265      & 0.130      & 0.532      & 2.77      &  --       &  --        \\
/OGLE-2011-BLG-012N &             & $\pm$0.001 & $\pm$0.0002 & $\pm$0.26 & $\pm$0.001 & $\pm$0.003 & $\pm$0.001 & $\pm$0.04 &  --       &  --        \\
                    & 5561.3/5558 & 5744.731   & 0.0140      & 17.12     & 4.597      & 0.200      & 0.514      & 3.12      &  --       &  --        \\
                    &             & $\pm$0.001 & $\pm$0.0002 & $\pm$0.21 & $\pm$0.027 & $\pm$0.005 & $\pm$0.001 & $\pm$0.04 &  --       &  --        
\enddata  
\tablecomments{ 
HJD'=HJD-2450000.
}
\end{deluxetable*}

In Table \ref{table:four}, we present the solutions of the parameters determined from modeling for the individual events. For events where the additional parallax modeling is conducted, we also list the parallax solution. 
In addition, we present the solutions of local minima if they exist. The model light curves resulting from the best fit solutions of the individual events are presented on 
the top of the observed data points in Figures 1 -- 7. In Figure \ref{fig:eight}, we also present the geometry of the lens system corresponding to the best fit solution of each event. Based on the best fit 
solutions, it is found that the mass ratios of the analyzed events are in the range $0.09 < q < 0.20$, indicating that the lower mass companions of the lenses are strong brown dwarf candidates.

For 3 out of the total 7 events, it is found that there exist local minima other than the best fit solution. The events with degenerate solutions include MOA-2011-BLG-104/OGLE-2011-BLG-0172, MOA-2011-BLG-149, and MOA-2011-BLG-278/OGLE-2011-BLG-012N. 
For all of these events, the degeneracy is caused by the symmetry of the lens equations between the close $(s<1)$ and wide $(s>1)$ binaries which is known as the ``close/wide binary degeneracy'' \citep{dominik99}. 
For MOA-2011-BLG-104/OGLE-2011-BLG-0172 and MOA-2011-BLG-149, the degeneracy is very severe with $\Delta\chi^2=2.5$ and $3.1$, respectively. For MOA-2011-BLG-278/OGLE-2011-BLG-012N, the degeneracy is less severe and the wide binary solution is 
preferred over the close binary solution with $\Delta\chi^2=259.4$. We note that the mass ratios of the pair of close and wide binary solutions are similar to each other although the binary separations are very different. Therefore, 
the degeneracy does not affect the brown dwarf candidacy of the lens.

For three events, it is possible to measure the lens mass by simultaneously measuring the lens parallax and the Einstein radius. With these values, the lens mass and distance are determined by
\begin{equation}
M = {\theta_{\rm E} \over{\kappa\pi_{\rm E}}},
\end{equation}
and
\begin{equation}
D_{\rm L} = {{\rm AU}\over{\pi_{\rm E}\theta_{\rm E}+\pi_{\rm S}}},
\end{equation}
respectively. Here $\pi_{\rm S}={\rm AU}/D_{\rm S}$ represents the parallax of the source star.

\begin{figure}[ht]
\epsscale{1.1}
\plotone{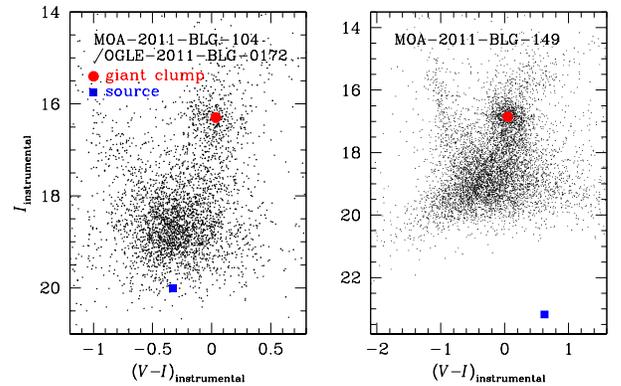}
\caption{\label{fig:nine}
Color-magnitude diagram of stars in the field of MOA-2011-BLG-104/OGLE-2011-BLG-0172 and MOA-2011-BLG-149
with the locations of the lensed stars and giant clumps marked as blue and red dots, respectively. 
}\end{figure}

For MOA-2011-BLG-104/OGLE-2011-BLG-0172, the value of the lens parallax 
measured from modeling is $\pi_{\rm E}=({\pi_{{\rm E},N}}^2+
{\pi_{{\rm E},E}}^2)^{1/2}=0.34\pm0.21$. The Einstein radius is measured 
based on the normalized source radius determined from modeling combined 
with the information of the angular source size by $\theta_{\rm E} = 
\theta_{\star}/\rho_{\star}$. The angular source size is assessed from the 
de-reddened color and brightness of the source star, which are estimated 
based on the source brightness measured in $V$ and $I$ pass bands and 
calibrated by the relative position of the source on the color-magnitude 
diagram with respect to a reference \citep{yoo04}. We use the centroid of 
bulge giant clump as the reference because its de-reddened brightness of 
$I_{0,{\rm c}}=14.45$ at the Galactocentric distance 
\citep{nataf12}\footnote{For other Galactic bulge fields with Galactic 
coordinates $(l,b)$, the I-band brightness is given by $I_{0,{\rm c}}=
14.45-5 \log (\cos l + \sin l\cos\phi/sin\phi)$, where $\phi\sim 45^\circ$ 
is the orientation angle of the Galactic bulge with respect to the line of 
sight.} and the color of $(V-I)_{0,{\rm c}}=1.06$ \citep{bensby11} are 
known.  We measure the de-reddened color and brightness of the source star 
to be $(V-I,I)_{0,{\rm s}}=(0.70,18.15)$.  Figure \ref{fig:nine} shows the 
locations of the source star and the centroid of the giant clump in the 
color-magnitude diagram that is constructed based on the CTIO data. After 
the initial photometric characterization of the source, we learned that 
spectroscopic observation is conducted when the source was highly magnified 
(Bensby 2012, private communication). From this spectroscopic observation, 
the de-reddened color of the source star is estimated as $(V-I)_{0,{\rm s}}
=0.62$, which is adopted for our analysis.  Once the de-reddened $V-I$ color 
of the source star is measured, it is translated into $V-K$ color by using 
the $V-I$ versus $V-K$ relations of \citet{bessell88} and then the source angular 
radius is estimated by using the relation between the $V-K$ color and the 
angular radius given by \citet{kervella04}. The measured angular source 
radius and the Einstein radius are $\theta_\star=(0.68\pm 0.06)\ \mu{\rm as}$ 
and $\theta_{\rm E}=(0.54\pm0.07)$ mas, respectively. With the measured 
Einstein radius, the mass and distance to the lens are estimated as $M=
(0.20\pm0.12)$ $M_{\odot}$ and $D_{\rm L}=(3.29\pm1.20)$ kpc, respectively. 
Then, the corresponding masses of the primary and the companion are $M_1=
M/(1 + q)= (0.18\pm0.11)\ M_{\odot}$ and $M_2=qM_1=(0.02\pm0.01)\ M_{\odot}$, 
respectively.  Therefore, the binary lens is composed of a low mass M dwarf 
and a brown dwarf.

For MOA-2011-BLG-149, the lens parallax is also measured with $\pi_{\rm E}
=0.78\pm0.04$. For this event, however, multi-band ($V$ and $I$) observations 
taken at a single observatory (CTIO) covered only near the high magnification peak 
of the light curve. As a result, the source brightness in both bands cannot 
be estimated by the usual method based on modeling. Fortunately, there exists 
no significant blending for this event and thus the color measured at the 
peak can be approximated as the source color. From the source brightness and 
magnification at the peak, we also estimate the 
baseline source brightness. The source location estimated in this way is 
plotted on the color-magnitude diagram presented in Figure \ref{fig:nine}. 
Once the source location is determined, the de-reddened color and brightness 
are determined by the usual way of using the clump centroid.  The measured 
values are $(V-I,I)_{0,{\rm s}} = (1.64, 20.8)$, which correspond to those of a Galactic 
bulge K-type dwarf with an angular radius of $\theta_\star=(0.46 \pm 0.04)\ 
\mu{\rm as}$.  Then, the estimated Einstein radius is $\theta_{\rm E}=(1.04
\pm0.10)$ mas and the resulting lens mass and distance are $M=(0.16\pm0.02)$ 
$M_{\odot}$ and $D_{\rm L}=(1.07\pm0.10)$ kpc, respectively.  The masses of 
the individual lens components are $M_1=(0.14\pm0.02)$ $M_{\odot}$ and $M_2
=(0.019\pm0.002)$ $M_{\odot}$, respectively.  Therefore, the lens system is 
also composed of a low mass M dwarf and a brown dwarf.

For another parallax event OGLE-2004-BLG-035, no multi-band observation 
is conducted. However, the source star of the event is very bright, 
implying that the source flux is not much affected by blended light. 
Under this approximation, we measure the source color in the OGLE catalog. 
Based on this approximate color combined with the giant clump location,
we estimate the de-reddened color and brightness of the source star of 
$(V-I,I)_{0,{\rm s}}=(1.19,15.83)$, which correspond to the values of a low luminosity 
giant star with a source radius of $\theta_{\star}= (3.72\pm0.32)$ $\mu$as. 
Considering that there exists a non-zero blended flux fraction $F_{\rm b}/
F_{\rm tot}\sim 20\%$, we decrease the source radius by $\sim (F_{\rm b}/
F_{\rm tot})/2 =10\%$.  The obtained Einstein radius is $\theta_{\rm E}=
(1.40\pm0.21)$ mas. Then, the mass of the lens system is 
$M=(2.48 \pm 0.87)\ M_{\odot}$ and thus the mass of the companion is 
$M_2=(0.29\pm0.10)\ M_{\odot}$. The companion mass is greater than the 
upper mass limit of brown dwarfs of $0.08$ $M_{\odot}$.

\section{CONCLUSION}

We searched for candidate binaries with brown dwarf companions by analyzing binary microlensing events discovered during the 2004 -- 2011 observing seasons. Under the low mass ratio criterion of 
$q < 0.2$, we found 7 candidate events, including OGLE-2004-BLG-035, OGLE-2004-BLG-039, OGLE-2007-BLG-006, OGLE-2007-BLG-399/MOA-2007-BLG-334, MOA-2011-BLG-104/OGLE-2011-BLG-0172, MOA-2011-BLG-149, 
and MOA-201-BLG-278/OGLE-2011-BLG-012N. Among them, we confirmed that the companions of MOA-2011-BLG-104/OGLE-2011-BLG-0172 and MOA-2011-BLG-149 were brown dwarfs by measuring the lens masses.

The number of microlensing brown dwarfs is expected to increase with the improvement of lensing surveys and analysis technique. With the adoption of a new wide field camera, the number of lensing events 
detected by the OGLE survey is increased by a factor more than 2. In addition, a new survey based on a network of 3 telescopes located at three different locations of the Southern Hemisphere 
(Korea Microlensing Telescope Network) is planned to operate from the 2014 season. Along with the improvement in the observational side, the analysis technique has greatly improved for the last several 
years, enabling prompt and precise analysis of a large number of anomalous events. With this improvement, the number of microlensing brown dwarfs will rapidly increase, making microlensing one of the major 
methods in the discovery of brown dwarfs.

\mbox{}

\acknowledgments 
Work by CH was supported by the Creative Research Initiative Program 
(2009-0081561) of National Research Foundation of Korea.
The OGLE project has received funding from the European Research Council 
under the European Community's Seventh Framework Programme 
(FP7/2007-2013) / ERC grant agreement no. 246678. The MOA experiment was 
supported by grants JSPS22403003 and JSPS23340064. 
TS was supported by the grant JSPS23340044.
Y. Muraki acknowledges support from JSPS grants JSPS23540339 and JSPS19340058.
The MiNDSTEp monitoring campaign is powered by ARTEMiS
(Automated Terrestrial Exoplanet Microlensing Search; Dominik et al.
2008, AN 329, 248). MH acknowledges support by the German Research
Foundation (DFG). DR (boursier FRIA), FF (boursier ARC) and J. Surdej acknowledge support
from the Communaut\'{e} fran\c{c}aise de Belgique Actions de recherche
concert\'{e}es -- Acad\'{e}mie universitaire Wallonie-Europe. KA, DMB,
MD, KH, MH, CL, CS, RAS, and YT are thankful to Qatar National Research
Fund (QNRF), member of Qatar Foundation, for support by grant NPRP
09-476-1-078.
CS received funding from the European Union Seventh Framework Programme (FPT/2007-2013) under grant agreement 268421.
This work is based in part on data collected by MiNDSTEp with the Danish 1.54 m telescope at the ESO La Silla Observatory. 
The Danish 1.54 m telescope is operated based on a grant from the Danish Natural Science Foundation (FNU). 
A. Gould and B.S. Gaudi acknowledge support from NSF AST-1103471.
B.S. Gaudi, A. Gould, and R.W. Pogge acknowledge support from NASA grant NNG04GL51G.
Work by J.C. Yee is supported by a National Science Foundation Graduate Research Fellowship under Grant No. 2009068160.
S. Dong's research was performed under contract with the California Institute of Technology (Caltech) funded by NASA through the Sagan Fellowship Program.
Research by TCH was carried out under the KRCF Young Scientist Research Fellowship Program. 
TCH and CUL acknowledge the support of Korea Astronomy and Space Science Institute (KASI) grant 2012-1-410-02.
IAB and PCMY acknowledge the support by the Marsden Fund of New Zealand.

\end{document}